\documentclass[trackchanges,twocolumn]{aastex701}

\usepackage{lmodern}
\usepackage{CJKutf8}
\usepackage{amsmath}
\usepackage{hyperref}
\usepackage{xspace}
\usepackage{caption}
\usepackage{upgreek}
\usepackage{ulem}

\begin{document}

\title{Fast X-ray Transient EP260321a: Shock Breakout through an Extended Circumstellar Matter}

\author[orcid=0000-0003-3031-6105]{Maokai Hu}
\affiliation{Physics Department, Tsinghua University, Beijing 100084, China}
\email[show]{kaihukaihu123@mail.tsinghua.edu.cn\\}  

\author[orcid=0000-0002-6535-8500]{Yi Yang\begin{CJK*}{UTF8}{gbsn}
(杨轶)\end{CJK*}} 
\affiliation{Physics Department, Tsinghua University, Beijing 100084, China}
\email[show]{yi\_yang@mail.tsinghua.edu.cn}


\author[0000-0002-7334-2357]{Xiaofeng Wang}
\affiliation{Physics Department, Tsinghua University, Beijing 100084, China}
\email{wang\_xf@mail.tsinghua.edu.cn}


\begin{abstract}

The Einstein Probe (EP) has discovered a fast X-ray transient, EP260321a, whose optical counterpart was subsequently classified as a broad-lined Type Ic supernova. 
Its observed rise time of about 690 seconds is inconsistent with theoretical predictions for shock breakout (SBO) from a Wolf-Rayet star, which is typically regarded as the progenitor of envelope-stripped supernovae. In this paper, we propose that EP260321a arose from a SBO within extended circumstellar matter (CSM), where breakout photons were initially trapped, then diffused out of the unshocked CSM as the ejecta expanded and the opacity decreased. We modeled this
photon diffusion process using both Monte Carlo simulations and an analytical approach to describe the temperature evolution. Through Markov Chain Monte Carlo sampling, we obtain the best-fit CSM parameters of ${\rm log}_{10}(M_{\rm csm}/M_{\odot}) = -3.3_{-0.2}^{+0.3}$ for the total CSM mass, $R_{\rm in} = 160_{-23}^{+45}\,R_{\odot}$ for the inner CSM radius, and $R_{\rm out} = 612_{-35}^{+47}\,R_{\odot}$ for the outer CSM radius. The presence of such a dense CSM indicates that the progenitor of EP260321a underwent episodic mass loss just days before the explosion, providing new observational constraints on the late-stage evolution of massive stars.

\end{abstract}



\section{Introduction}

Shock breakout (SBO) occurs when the radiation field carries energy ahead of the shock front reaching the stellar surface \citep{1978ApJ...223L.109K,1992ApJ...393..742E,1999ApJ...510..379M,2010ApJ...725..904N,2012ApJ...747..147K,2020MNRAS.494.3927K}. Instead of particle collisions, the radiation carries energy ahead of the shock, preheating and accelerating the upstream gas. This establishes a radiation-mediated shock. As the shock reaches the star's surface, it heats the material to extremely high temperatures, emitting high-energy photons predominantly in X-ray and ultraviolet wavelengths, and producing the earliest electromagnetic signal of a core-collapse supernova (SN). Such prompt emission carries key information encoding the progenitor structure and explosion mechanism. Thus far, several X-ray transients in the XMM-Newton archive have been proposed as SBO candidates \citep{2020ApJ...896...39A,2020ApJ...898...37N}, while XRF\,080109, which has been serendipitously discovered by the {\it SWIFT} X-ray Telescope, remains the only X-ray SBO event unambiguously associated with an SN \citep{2008Natur.453..469S,2008Sci...321.1185M}. Its optical counterpart, SN~2008D, was initially identified as a broad-lined Type Ic (Ic-BL) SN but was later reclassified as a Type Ib SN following the appearance of prominent helium absorption lines \citep{2008ApJ...683L.135C,2009ApJ...692.1131T,2009ApJ...692L..84M}. 

The Einstein Probe (EP;~\citealp{2015arXiv150607735Y, 2022hxga.book...86Y}) is a space mission dedicated to high-sensitivity, all-sky monitoring in the soft X-ray band to detect cosmic transients and explosive astrophysical phenomena. The Wide-field X-ray Telescope (WXT) onboard the EP utilizes advanced lobster-eye micro-pore optics, achieving a single-exposure field of view of about 3600 square degrees \citep{2022hxga.book...86Y}. During its first two years of operation, EP detected over 100 Fast X-ray Transients (FXTs), some of which have been confirmed to be associated with SNe~Ic-BL \citep{2025ApJ...988L..14E, 2025ApJ...988L..34J, 2025arXiv250417034L, 2025ApJ...982L..47V, 2025NatAs...9.1073S, 2025ApJ...988L..13R, 2026ApJ..1006L..36S, 2026arXiv260606213C, 2026ApJ...996L..29V}. Compared to the prompt emission of gamma-ray bursts (GRBs), these FXTs exhibit softer spectra, longer durations, and substantially lower total energy. The physical origin of FXTs remains uncertain; prevailing hypotheses include jet-cocoon systems, refreshed shocks in GRBs, post-merger magnetars, or SBO emission from the progenitor \citep{2025ApJ...986L...4H, 2025ApJ...978L..21S, 2025ApJ...988L..60S, 2025ApJ...985...21Z, 2026ApJ..1005..214P}. 

EP260321a is an intriguing FXT discovered by EP \citep{2026GCN.44068....1H,2026GCN.44075....1H}. With an onset time of $T_{0}=$2026-03-21 12:16:08 UTC, or MJD 61120.511~\citep{2026arXiv260610014Y}, the transient displays a rise time of $\approx$690\,s and a radiation temperature of about 0.1 keV. Motivated by the prompt optical follow-up observations starting from as early as $\sim T_{0}+1$\,hr, which identified a rapidly brightening source~\citep{2026GCN.44070....1L,2026GCN.44082....1T,2026GCN.44089....1S,2026GCN.44103....1P}, subsequent spectroscopic observations confirmed that the optical counterpart of EP260321a is the Ic-BL SN~2026gzf \citep{2026ApJ..1006L..13O, 2026ApJ..1007L...9R, 2026arXiv260610009C, 2026arXiv260610002M, 2026arXiv260610014Y}. This association indicates that the progenitor may have been a Wolf-Rayet (WR) star, whose outer layers of hydrogen and helium are stripped away. However, the observed rise time ($\sim 690\,$s) is approximately an order of magnitude greater than the theoretical expectation for SBO from a bare WR star \citep{2010ApJ...725..904N}. This discrepancy points to the presence of circumstellar matter (CSM), through which the SBO must propagate before its X-ray emission can escape \citep{2026ApJ..1008L..63D,2026ApJ..1007L...9R,2026arXiv260724045S,2026arXiv260610014Y}.  

Previous studies employed a scenario in which the SBO emerges from a compact CSM to explain the luminosity and temperature evolution of EP260321a \citep{2026ApJ..1007L...9R, 2026arXiv260610014Y}. Here, we suggest that their assumed compact-CSM framework may be internally inconsistent with the parameters they derived. In the compact-CSM regime, the CSM is optically thick that SBO occurs at the outer edge of the shell. A representative CSM radius ($\sim 3\times10^{12}$ cm) and CSM mass ($\sim 1\times10^{-6}\ M_\odot$) reported in \cite{2026ApJ..1007L...9R} indicate an optical depth of $\sim 2.0$, for which we assume the optical opacity ($\kappa$) of 0.2 g$\,$cm$^{-3}$ corresponding to hydrogen-poor gas. These CSM parameters therefore imply an optically thin regime, where a well-defined SBO cannot occur. The CSM parameters reported in \cite{2026arXiv260610014Y} appear similarly inconsistent. With an inner radius of $\sim 260\ R_\odot$, an outer radius of $\sim 320\ R_\odot$, and a total CSM mass of $\sim 10^{-4}\ M_\odot$, the total optical depth is estimated to be $\sim 10$. Such a value is insufficient to establish an optically thick regime, suggesting that the SBO is unlikely to occur near the outer edge of the CSM. 

We instead propose an extended CSM scenario, in which the SBO emerges within the CSM shell, and the resulting photons diffuse through the unshocked CSM before reaching the observer. In Section~\ref{Sec_II}, we describe the physical framework of SBO from the extended CSM. Section~\ref{Sec_III} presents the analysis and fitting results under this framework. The discussions and conclusions are provided in Section~\ref{Sec_IV}.

\section{Method} 
\label{Sec_II}

\begin{figure*}[ht!]
\plotone{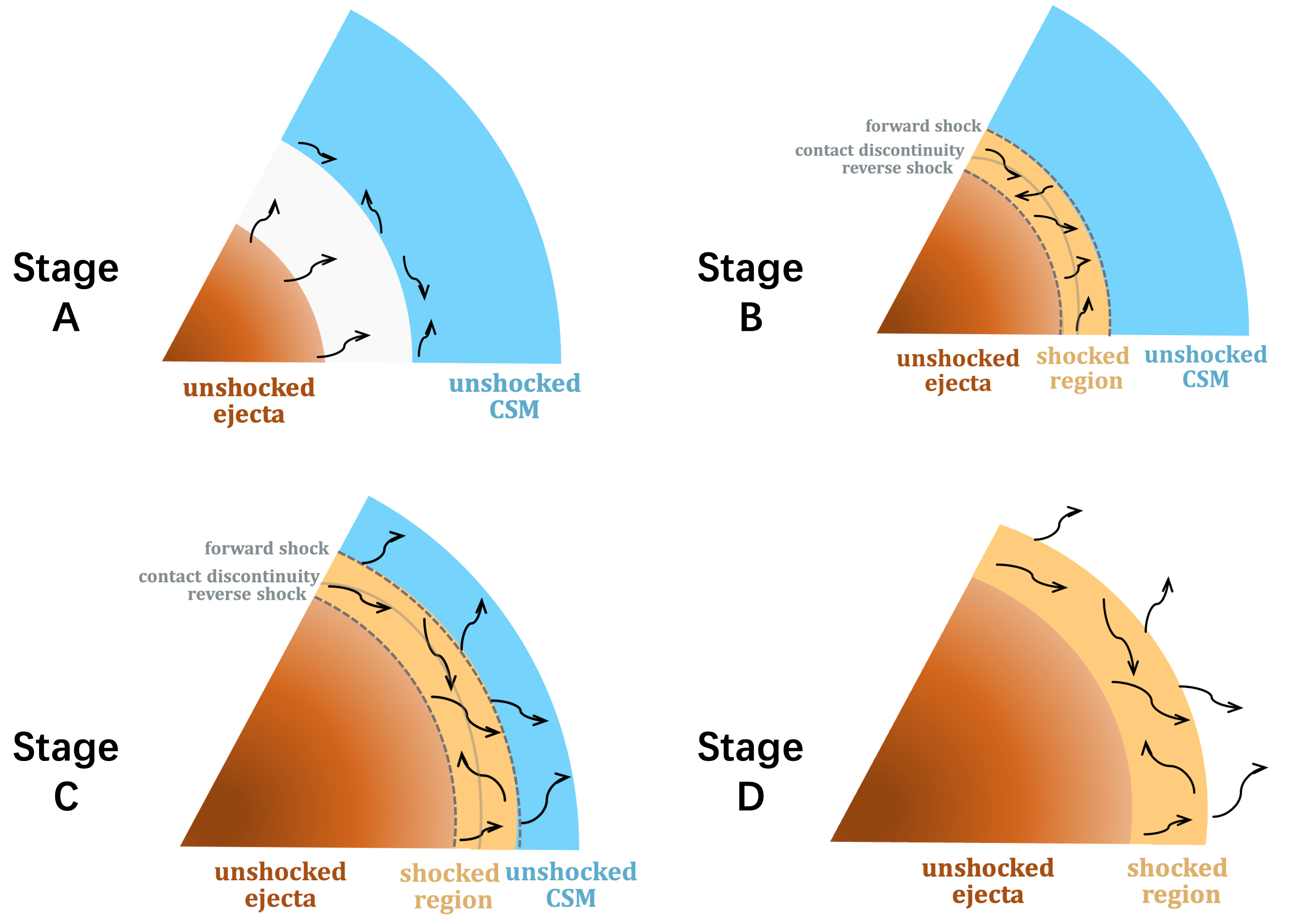}
\caption{Schematic illustration of the four evolutionary stages of shock interaction with an extended CSM shell. Stage A: ionization of the CSM by radiation from the SBO at the progenitor's surface. Stage B: radiative energy accumulation, in which a forward shock propagates through the optically thick CSM and radiation is trapped behind the shock front. Stage C: SBO of the extended CSM while photons diffuse outward through the unshocked CSM. Stage D: cooling phase after the shock has swept up the entire CSM shell.
\label{fig1_shiyi}}
\end{figure*}

\begin{figure*}[ht!]
\plotone{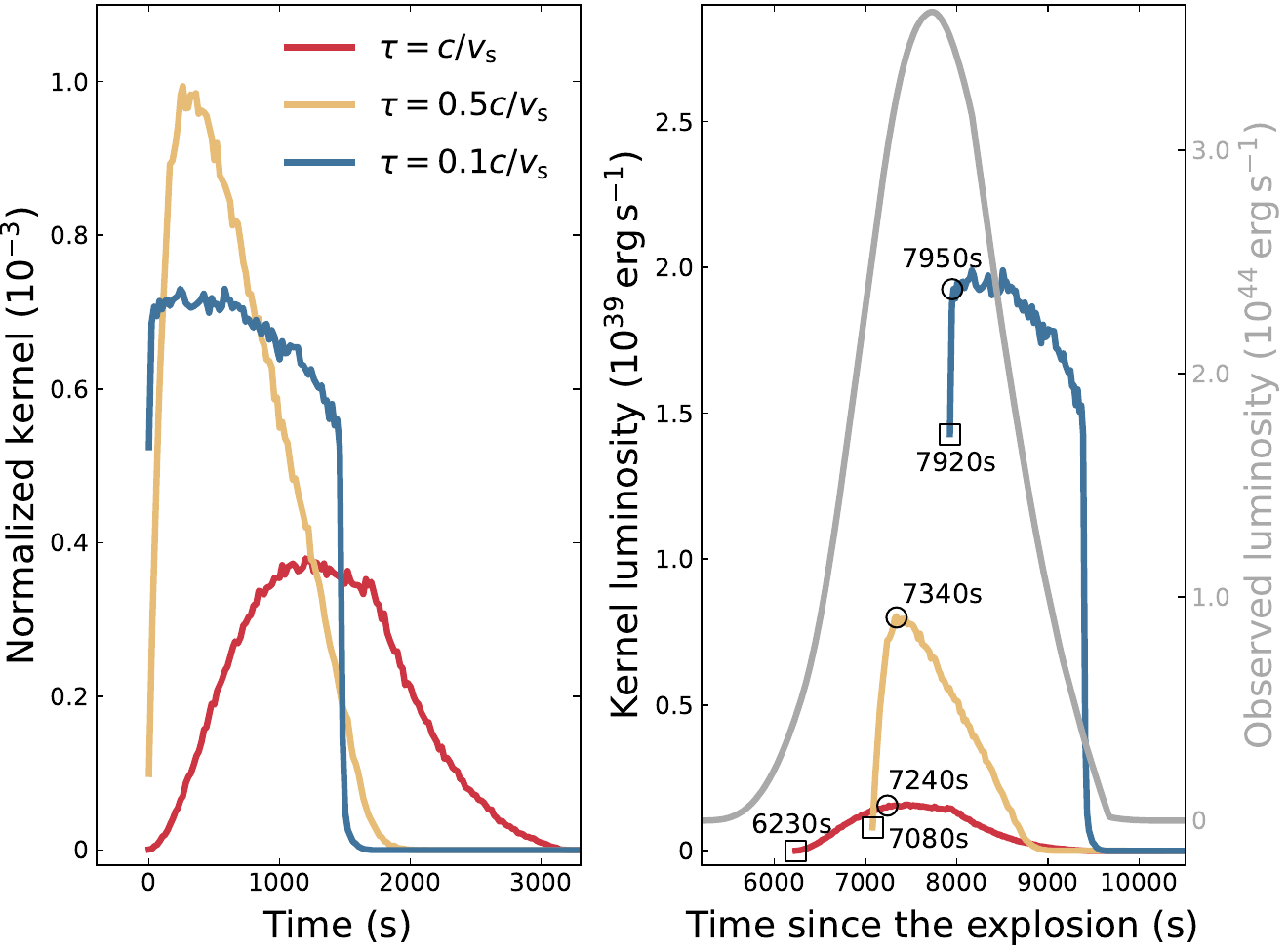}
\caption{For illustration purposes, only three diffusion kernels are displayed corresponding to the optical depths at which photons are injected, namely $\tau = c/v_{\rm s}$ (red line), $\tau = 0.5c/v_{\rm s}$ (yellow line), and $\tau = 0.1c/v_{\rm s}$ (blue line), respectively. The left panel presents the kernel profiles, which have been normalized so that each integrates to unity. The right panel displays the observed light curves resulting from photons diffusing out of the CSM at the three corresponding optical depths. 
The black squares mark when photons are injected into the unshocked CSM at each optical depth, while the black circles indicate when the corresponding light curves reach their peak. The gray line shows the observed light curve, synthesized by integrating the contributions from all kernels as the shock propagates through the outer boundary of the CSM.
\label{fig2_kernel}}
\end{figure*}

Massive stars may experience episodic mass loss during the final stages of their evolution, leaving CSM in the vicinity of the SN \citep{2012MNRAS.423L..92Q,2014ApJ...789..104O,2014Natur.509..471G,2014ARA&A..52..487S,2016ApJ...818....3K,2017NatPh..13..510Y,2017A&A...605A..83D,2021ApJ...912...46B,2023ApJ...952..119B,2024ApJ...970..189J}. When the SN shock propagates into this CSM, the resulting observational signatures depend critically on the total optical depth of the CSM, $\tau$ \citep{2011ApJ...729L...6C,2012ApJ...746..121C,2013ApJ...763...42O,2014ApJ...788..113S,2017A&A...605A..83D,2017ApJ...835..140M,2017MNRAS.469L.108M,2017hsn..book..875C,2022ApJ...928..122M}. We can distinguish three regimes based on the ratio of $\tau$ to $c/v_{\rm s}$, where $c$ is the speed of light and $v_{\rm s}$ is the shock velocity: 

\textbf{Optically thin regime ($\tau < c/v_{\rm s}$):} Radiation generated by ejecta-CSM interaction is injected into the unshocked CSM without being trapped behind the forward shock (e.g., \citealt{2013MNRAS.435.1520M}). In this regime, a well-defined SBO does not occur, and hard, non-thermal X-rays dominate the emission (e.g., \citealt{2022ApJ...928..122M}). Previous analyses of EP260321a inferred that the prompt X-ray emission was driven by interaction between the SN ejecta and an optically thick CSM, manifesting a pure thermal spectrum~\citep{2026arXiv260610014Y}. Therefore, the optically thin CSM scenario is disfavored in our study.

\textbf{Compact-CSM regime ($\tau \gg c/v_{\rm s}$):} The 'compact CSM' refers to the case in which the SBO emerges near the outer boundary of the CSM \citep{2021ApJ...910..128H}. This scenario is characterized by a sufficiently high CSM optical depth such that the location where $\tau \sim c/v_{\rm s}$ is close to the outer edge of the CSM. When $\tau$ is large, photons trapped behind the shock front typically achieve full thermalization, resulting in SBO radiation with a blackbody spectral shape \citep{2022ApJ...933..238M}. We find that the compact-CSM model does not fully match the observational signatures of EP260321a. 

\textbf{Extended CSM regime ($\tau \sim {\rm a\ few} \times c/v_{\rm s}$):} At the moment the shock emerges from the stellar surface and enters the immediate CSM, the unshocked CSM remains optically thick. This dense medium traps the radiation behind the shock and drives it toward thermalization. SBO occurs when the shock reaches the location of optical depth $\tau\sim c/v_{\rm s}$. Because the shock remains embedded in the CSM at breakout, shock-generated photons reach the observer only after diffusing through the unshocked CSM. This diffusion process within the unshocked CSM reshapes the light curve relative to the intrinsic shock luminosity. Unlike in the compact-CSM regime, the SBO radiation here deviates moderately from thermal equilibrium. 

In the following subsections, we outline the basic physical framework of SBO through an extended CSM shell and discuss the temperature evolution and photon diffusion process in detail.

\subsection{Physical Picture of SBO in Extended CSM}    
\label{Sec_2.1}

SBO in extended CSM involves two primary physical components: the dynamics of shock propagation through the CSM, and the radiative processes governing thermalization and diffusion. The dynamical framework has been well established in previous studies (e.g., \citealt{1994ApJ...420..268C,2004ApJ...616..339W,2006ApJ...651..381C,2012ApJ...747L..17C,2012ApJ...759..108S,2020PASJ...72...67T,2022ApJ...934....5Y,2023MNRAS.525..246H,2025ApJ...978L..27H,2025ApJ...984...44H}), where the time evolution of the shock is described analytically or numerically in terms of key parameters including the energy and mass of the ejecta ($E_{\rm ej}, M_{\rm ej}$), the CSM mass-loss rate ($\dot{M}_{\rm w}$), and the wind velocity ($v_{\rm w}$). We model the CSM as a detached spherical shell characterized by a well-defined inner radius $R_{\rm in}$, outer radius $R_{\rm out}$, and total mass $M_{\rm csm}$. This configuration corresponds to a mass-loss rate of $\dot{M}_{\rm w}/v_{\rm w} = M_{\rm csm}/(R_{\rm out}-R_{\rm in})$ assuming a CSM density profile of $\rho_{\rm csm} = \frac{\dot{M}_{\rm w}}{v_{\rm w}4\pi R^2}$, where $R$ is the distance from the SN. The photon-diffusion process during shock interaction with a radially extended CSM can be divided into four sequential stages, as illustrated in Figure~\ref{fig1_shiyi}: 

\textbf{Stage A:} Before the SN ejecta collides with the CSM, radiation from the SBO at the progenitor's surface is injected into the CSM, ionizing and diffusing through the surrounding gas. The internal energy of the SBO from a typical WR star yields $E_{\rm wr} = 9\times10^{44}\,{\rm erg}\, M^{-0.41}_{15}R^{1.66}_{5}E^{0.58}_{51}$ \citep{2010ApJ...725..904N}, where $M_{15}$, $R_{5}$, and $E_{51}$ represent the stellar mass, stellar radius, and explosion energy in units of $15\,M_{\odot}$, $5\,R_{\odot}$, and $10^{51}\,{\rm erg}$, respectively. This early CSM ionization is the defining characteristic of Stage A.
The remaining radiative energy is trapped within the optically thick CSM and eventually escapes the shell as the shock propagates outward. 

\textbf{Stage B:} The SN ejecta collide with the inner edge of the CSM, driving a forward shock through the CSM and a reverse shock into the ejecta. The high opacity of the unshocked CSM causes radiative energy to accumulate in the shocked region, where photons are thermalized through processes such as free-free emission and Compton scattering. This trapping and thermalization define Stage B. Consequently, emission produced at the SBO in the extended CSM is expected to emerge predominantly in soft X-rays rather than in the hard X-ray regime. Additionally, we utilize the numerical code developed in \cite{2023MNRAS.525..246H} to calculate the dynamical evolution of the shock propagating through the extended CSM. Further details of the thermalization process are provided in the following subsection.

\textbf{Stage C:} SBO occurs within the CSM shell once the optical depth of the unshocked CSM drops to an order of $c/v_{\rm s}$. Photons previously trapped behind the shock front are then released into the unshocked CSM and diffuse outward through electron scattering. This is the defining characteristic of Stage C. As the shock advances toward the edge of the CSM, the optical depth of the unshocked CSM ahead of it decreases rapidly. Therefore, photons injected at earlier epochs encounter longer diffusion timescales because the optical depth is higher at those stages. As a result, photons injected at later phases would escape the CSM in progressively shorter times, producing fast-evolving luminosity curves. We model the diffusion process using a Monte Carlo method developed by \cite{2025ApJ...984...44H}. 

\textbf{Stage D:} Cooling phase after the shock sweeps up the CSM. The emission shifts to optical and ultraviolet wavelengths as the shocked gas expands and cools. In this paper, we focus on the X-ray emission during Stages B and C. Detailed modeling of the photometric behavior during Stage D is beyond the scope of this work. 

\subsection{Radiation Temperature before the Shock Breakout} 
\label{Sec_2.2}

Radiation trapped behind the shock front is initially out of thermal equilibrium and is thermalized through free-free emission and Comptonization. According to \cite{2010ApJ...725..904N}, the radiation temperature $T$ is expressed in terms of the thermal coupling coefficient $\eta$ and the Comptonization parameter $\xi$ as $T = T_{\rm BB} \eta^2/\xi^2$ for $\eta >1$, and $T = T_{\rm BB}$ for $\eta < 1$, where $T_{\rm BB}$ is the thermal equilibrium temperature. Note that both $\eta$ and $\xi$ depend on the gas density $\rho_{\rm sh}$ and the photon energy density $\epsilon_{\rm sh}$; therefore, the evolution of the radiation temperature during Stage B can be determined by first calculating these two quantities.

We compute the average mass density of the shocked region as $\rho_{\rm sh} = M_{\rm sh} / V_{\rm sh}$, where $M_{\rm sh}$ and $V_{\rm sh}$ are the mass and volume of the shocked region, respectively. This simplification neglects the detailed density profile within the shocked region, which may introduce systematic deviations in the computed temperature evolution.

During Stage B, the forward shock continuously converts the ejecta's kinetic energy into radiative energy. The instantaneous shock luminosity is given by  $L = \frac{1}{2}\frac{\dot{M}_{\rm w}}{v_{\rm w}}v_{\rm s}^3$, where we assume an energy conversion efficiency of 1.0 and a fixed wind velocity of $v_{\rm w} = 1000\,{\rm km}\,{\rm s}^{-1}$ throughout this work. The total accumulated radiative energy behind the shock evolves as $E(t) = E_0 + \int L(t)\mathrm{d}t$, where $E_0$ is the trapped radiative energy from the SBO of the progenitor star. Consequently, the photon energy density and the thermal equilibrium temperature are expressed as $\epsilon_{\rm sh} = E(t) / V_{\rm sh}$, and $T_{\rm BB} = (\epsilon_{\rm sh} / a)^{1/4}$, respectively, where $a$ is the radiation constant. We then employ the framework outlined in \cite{2010ApJ...725..904N} to determine the radiation temperature and use a Wien-approximated spectral profile to calculate the X-ray luminosity in the 0.4-2.0 keV band. This enables a comparison with the published data of EP260321a in \cite{2026arXiv260610014Y}.

\begin{figure*}[ht!]
\plotone{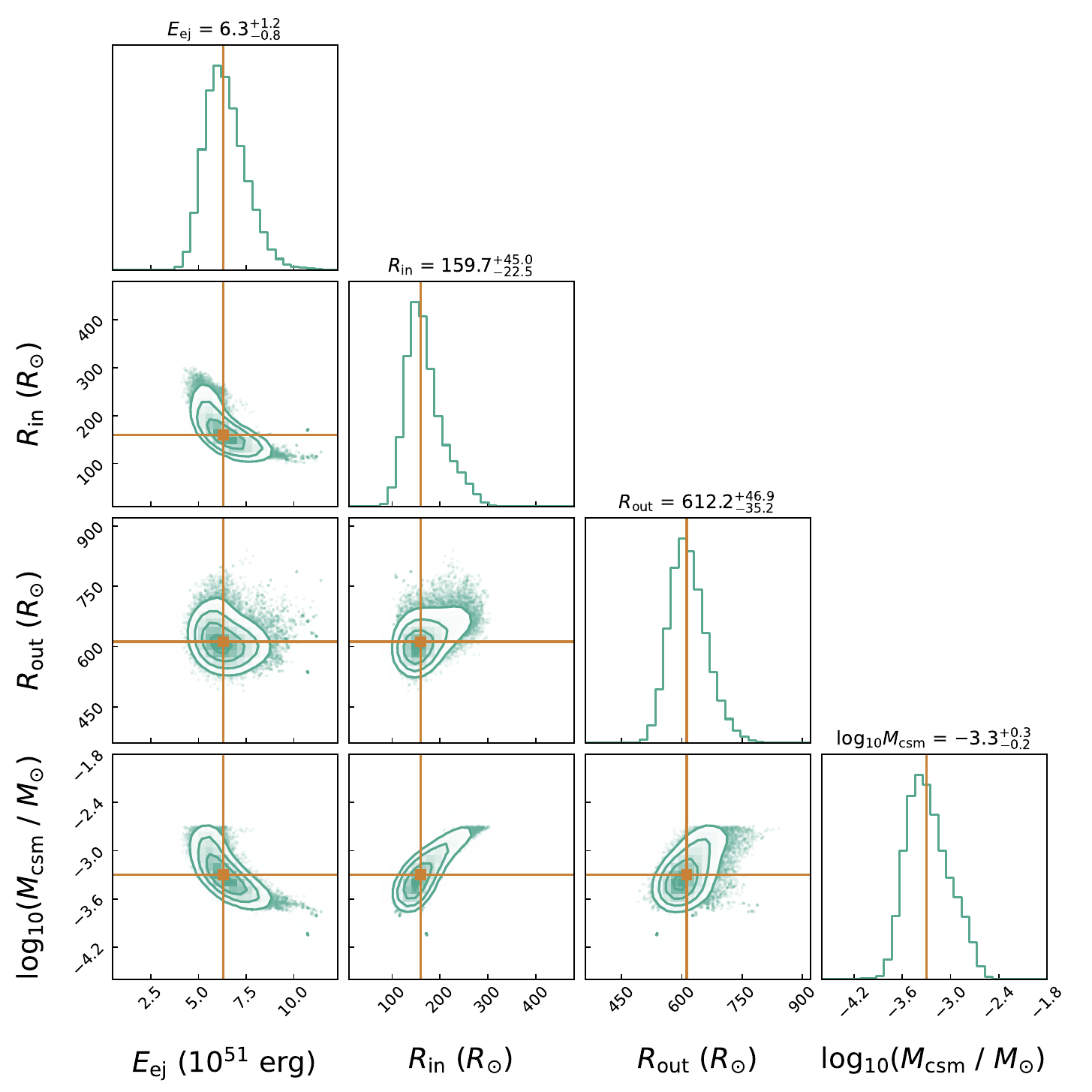}
\caption{Corner plot showing the posterior distributions of the four model parameters from MCMC sampling. The diagonal panels show the 1D marginalized posterior for each parameter, while the off-diagonal panels show 2D joint confidence contours. Solid lines in each panel mark the posterior median values as indicated by the text above each column. 
\label{fig3_mcmc}}
\end{figure*}

\begin{figure*}[ht!]
\plotone{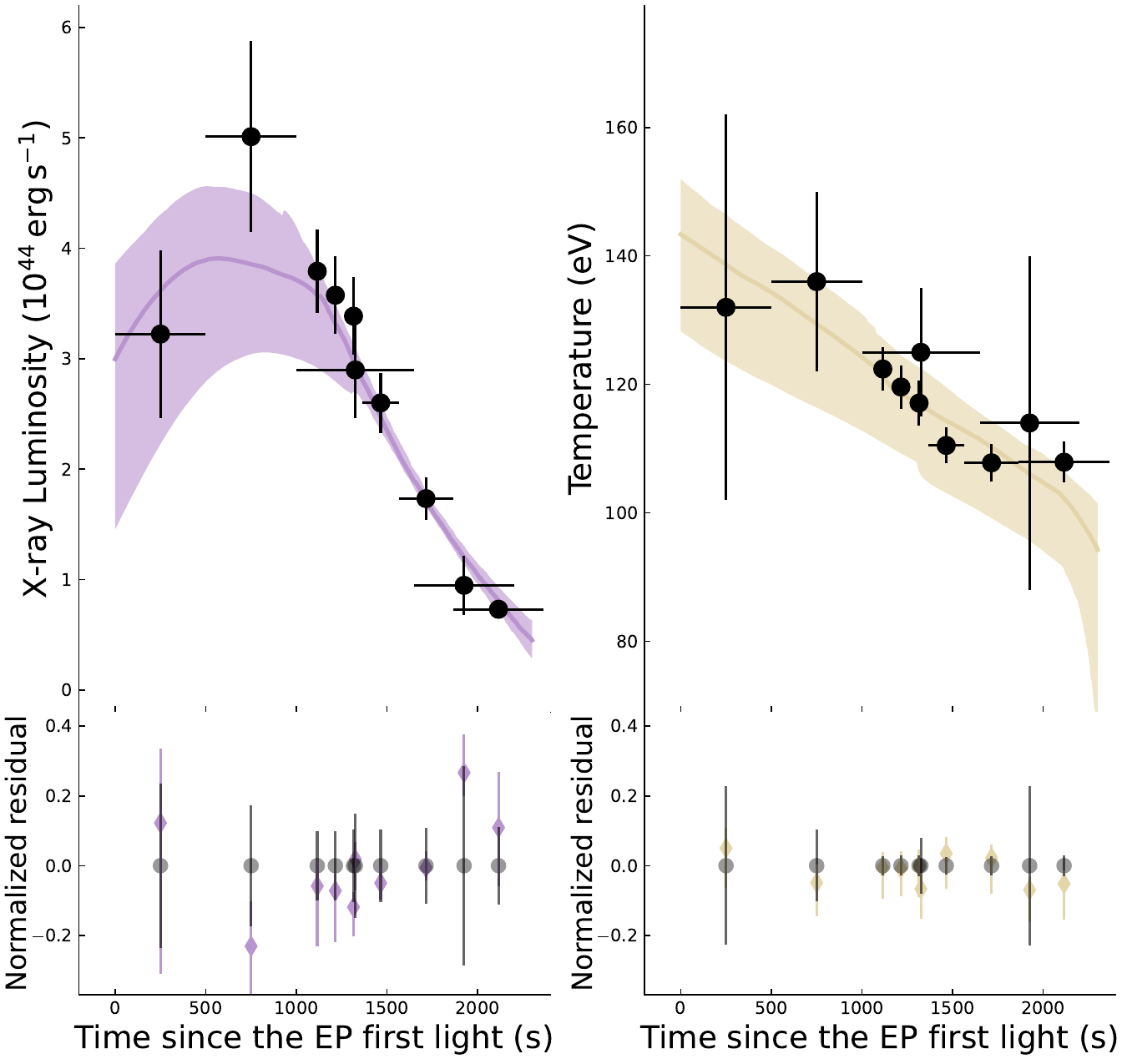}
\caption{Best-fit extended CSM model compared to the observations of EP260321a. Upper left: X-ray luminosity curve in the 0.4–2.0 keV band. Upper right: time evolution of the radiation temperature. Filled black circles present the X-ray luminosity and temperature adopted from \cite{2026arXiv260610014Y}. Phases are given relative to the first detection by EP at $T_{0}=$2026-03-21 12:16:08 UTC, or MJD 61120.511~\citep{2026arXiv260610014Y}. Solid lines show the best-fit model predictions, with the parameters $E_{\rm ej} = 7.0\times10^{51}\,{\rm erg}$, $R_{\rm in} = 141\,R_{\odot}$, $R_{\rm out}=580\,R_{\odot}$, and ${\rm log}_{10}(M_{\rm csm}/M_{\odot}) = -3.4$. The purple and yellow shaded regions indicate the 68\% posterior predictive ranges. The lower panels show the normalized residuals. The purple and yellow diamonds represent the normalized residuals of the model fit, while the error bars on the gray circles indicate the uncertainties of the data themselves.
\label{fig4_LC}}
\end{figure*}

\subsection{Photon Diffusion within the Unshocked CSM}
\label{Sec_2.3}

During Stage C, photons injected from the shocked region diffuse through the unshocked CSM via electron scattering, which conserves both photon number and energy. This diffusion process is modeled using a Monte Carlo code described in \cite{2025ApJ...984...44H}, which self-consistently tracks three key processes: photon injection from the shocked region, multiple electron scatterings within the unshocked CSM, and the collection of photons as they escape the outer boundary of the CSM shell.

The Monte Carlo simulation is initiated when the shock reaches an optical depth of $\tau \sim 2c/v_{\rm s}$ in the CSM, slightly before the formal SBO condition of $\tau \sim c/v_{\rm s}$. This approach enables accurate modeling of photon leakage prior to SBO. At this stage, the unshocked CSM ahead of the shock remains optically thick, and injected photons are predominantly back-scattered and largely confined within the shocked region. As the shock propagates outward, the optical depth of the unshocked CSM decreases rapidly, allowing an increasing fraction of photons to escape. Photon pulses injected at different shock positions, corresponding to different optical depths, show distinct escape-time distributions, which we refer to as diffusion kernels. 

Figure~\ref{fig2_kernel} shows the diffusion kernels computed fo $E_{\rm ej} = 5\times10^{51}\,{\rm erg}$, $R_{\rm in} = 250\,R_{\odot}$, $R_{\rm out} = 650\,R_{\odot}$, and $M_{\rm csm} = 1.5\times10^{-3}\,M_{\odot}$. The left panel displays the diffusion kernels for photon pulses injected at three representative optical depths: $\tau = c/v_{\rm s}$, $\tau = 0.5c/v_{\rm s}$, and $\tau = 0.1c/v_{\rm s}$. For $\tau = c/v_{\rm s}$, a fraction of photons are back-scattered and subsequently re-trapped behind the forward shock, resulting in an integrated fraction of escaped photons less than unity. Meanwhile, photons injected near $\tau \sim c/v_{\rm s}$ also undergo multiple scattering within the unshocked CSM, producing a broad diffusion-kernel distribution. At $\tau \sim 0.1c/v_{\rm s}$, the shock approaches the outer boundary of the CSM, allowing injected photons to escape with little to no scattering. This leads to a diffusion kernel that closely resembles a pure light-travel-time delay. 

The right panel of Figure~\ref{fig2_kernel} illustrates that photons injected into the unshocked CSM at different optical depths escape over compressed time intervals, producing a rapidly evolving integrated luminosity. The photon injection times corresponding to the three optical depths are $6230\,{\rm s}$, $7080\,{\rm s}$, and $7920\,{\rm s}$, while the times at which each diffusion kernel approaches its peak are $7240\,{\rm s}$, $7340\,{\rm s}$, and $7950\,{\rm s}$, respectively. The intervals between the kernel peak times are notably shorter than those between the injection times. As a result, the integrated luminosity evolves over a compressed timescale relative to the intrinsic shock luminosity, which does not account for diffusion through the unshocked CSM. Therefore, photon diffusion through a radially extended CSM can produce a rapidly-evolving light curve, as shown by the gray line in Figure~\ref{fig2_kernel}.

\section{Results}
\label{Sec_III}

\subsection{The Configuration of MCMC Fitting}

We employ Markov Chain Monte Carlo (MCMC) sampling to constrain the model parameters by fitting the observed X-ray light curve and temperature evolution of EP260321a. Our model has four free parameters: the ejecta kinetic energy $E_{\rm ej}$, the inner radius of the CSM shell $R_{\rm in}$, the radial width of the CSM $R_{\rm wid} = R_{\rm out} - R_{\rm in}$, and the total mass of the CSM $M_{\rm csm}$ in units of $10^{51}\,$ erg, $R_{\odot}$, $R_{\odot}$, and $M_{\odot}$, respectively. We adopt flat priors over the following ranges: $E_{\rm ej} \in (0.2, 20.0)$, $R_{\rm in} \in (100, 500)$, $R_{\rm wid} \in (50, 1000)$, and $M_{\rm csm} \in (10^{-4}, 5\times10^{-3})$. 

Note that the parameter pairs $(R_{\rm in}, R_{\rm wid})$ and $(R_{\rm in}, R_{\rm out})$ provide equivalent descriptions. However, treating $R_{\rm out}$ as a free parameter in the MCMC fit can yield unphysical solutions with $R_{\rm wid} < 0$. Using \(R_{\rm wid}\) as the free parameter avoids this issue. We therefore adopt $R_{\rm wid}$ as a free parameter in the MCMC fit and report the corresponding $R_{\rm out}$ results in Figure~\ref{fig3_mcmc}. The $R_{\rm out}$ results are not obtained by simply adding the median values of $R_{\rm in}$ and $R_{\rm wid}$. Instead, we first derive a set of $R_{\rm out}$ values by calculating $R_{\rm out} = R_{\rm in} + R_{\rm wid}$ for each MCMC-chain sample and derive the median and confidence intervals from the resulting posterior distribution. 

In our MCMC fitting, the likelihood function is expressed as $-\frac{1}{2}\sum_{n=1}^{N}\left[ \frac{(y_{\rm model, n} - y_{\rm data, n})^2}{s_{n}^2} + \ln(s_{n}^2)\right]$, where $y_{\rm model}$ and $y_{\rm data}$ represent the predicted and observed X-ray luminosities or radiation temperature, respectively. The dataset contains $N$ ($=20$) points: 10 X-ray luminosity measurements and 10 radiation-temperature measurements from Table~2 of \cite{2026arXiv260610014Y}. The variance is defined as $s^2 = \sigma_{\rm data}^2 + y_{\rm model}^2 \ln(2f)$, where $\sigma_{\rm data}$ is the 1-$\sigma$ uncertainty of the $n-$th data point. We set $f$ as a constant model parameter.

\subsection{Best-fit Parameters for the Extended CSM Model}

Using the parallel configuration of the Python package emcee \citep{2013PASP..125..306F}, we ran 2,000 MCMC steps with 200 walkers.   
Figure~\ref{fig3_mcmc} shows the resulting posterior parameter distributions as a corner plot. The posterior medians and $68\%$ confidence intervals are $E_{\rm ej} = 6.3^{+1.2}_{-0.8} \times10^{51}\,{\rm erg}$, $R_{\rm in} = 159.7^{+45.0}_{-22.5}\,R_{\odot}$, $R_{\rm out}=612.2^{+46.9}_{-35.2}\,R_{\odot}$, and ${\rm log}_{10}(M_{\rm csm}/M_{\odot}) = -3.3^{+0.3}_{-0.2}$. To identify the best-fit model, we drew 200 parameter sets randomly from within the $68\%$ confidence interval of the MCMC posterior distribution, from which we identified the best-fit model as $E_{\rm ej} = 7.0\times10^{51}\,{\rm erg}$, $R_{\rm in} = 141\,R_{\odot}$, $R_{\rm out}=580\,R_{\odot}$, and ${\rm log}_{10}(M_{\rm csm}/M_{\odot}) = -3.4$. These best-fit parameters lie close to the posterior medians, and the $\chi^2$ values of 13 and 15 indicate that the MCMC procedure has reasonably constrained the model parameters to the observational data. Figure~\ref {fig4_LC} shows the resulting fits to the X-ray light curve and temperature evolution. The model successfully reproduces both the rise and decay of the observed light curve and the overall temperature evolution. The observed duration is naturally explained by photon diffusion through the extended, unshocked CSM layer. 

The MCMC result suggests that the CSM shell extends from $\sim$150 to $\sim$600 $R_{\odot}$, corresponding to a radial width of $\sim$450 $R_\odot$. Assuming a wind velocity of 1000\,km\,s$^{-1}$, this extent implies that the mass-loss episode occurred approximately 1--5 days before the SN explosion. The inferred total CSM mass of $\sim10^{-3.3}\ M_\odot$ corresponds to a mass-loss rate of $\sim 0.05\ M_\odot\ \rm yr^{-1}$ during the eruption, several orders of magnitude above the steady-state rates of $10^{-5.6}-10^{-4.4}$\,M$_{\odot}$\,yr$^{-1}$ derived for Galactic nitrogen- and carbon-sequence WR stars (see, e.g.,~\citealp{2007ARA&A..45..177C}). 

The optical depth of the CSM corresponding to the best-fit model is $\tau \approx 30$, with a posterior median of $\tau = 34^{+17}_{-10}$, consistent with the condition for SBO occurring within the extended CSM ($\tau \sim {\rm a\ few}\times c/v_{\rm s}$).
However, the model adopts several simplifications that could potentially introduce systematic bias into the fitting process.
\begin{itemize}
\item It assumes uniform density and temperature throughout the shocked region, whereas in reality the shocked layer may exhibit structured density and temperature gradients. This approximation may introduce systematic uncertainties into the derived temperature evolution. 
\item We assume a steady-wind density profile for the unshocked CSM, $\rho_{\rm csm} \propto R^{-2}$ with no temporal variation in the mass-loss rate prior to the explosion. Departure from this profile would alter the photon diffusion timescale through the unshocked CSM and, consequently, the inferred radial extent and radial distribution of CSM.
\item Our calculation of the photon diffusion in the unshocked CSM accounts solely for electron-scattering opacity and neglects other processes such as free-free emission and Comptonization. 
\item The model assumes full ionization of all elements, leaving out line absorption and emission effects, which may alter the spectral and light-curve characteristics. 
\item We assume that the conversion efficiency from kinetic energy to internal energy is unity throughout the ejecta-CSM interaction, consistent with previous studies \citep{2010ApJ...724.1396O,2021ApJ...910..128H,2022ApJ...933..238M,2026ApJ...998..247W}. While this is a reasonable approximation at early phases when the optical depth is high, the efficiency is expected to decline as the shock propagates outward into lower-density material. This simplification may therefore cause the model to overestimate the predicted luminosity at later epochs.
\item We assume a spherical CSM geometry, whereas asymmetric CSM and ejecta, as well as an axisymmetric ionization structure, were inferred for EP260321a \citep{2026ApJ..1008L...3W}. Moreover, an axisymmetric SBO has also been inferred for core-collapse SNe~\citep{2025SciA...11x2925Y}. 
\end{itemize}  
Nonetheless, the model provides a physically self-consistent interpretation of EP260321a and effectively captures the primary physics of SBO through an extended CSM.

\section{Discussion and Conclusion} 
\label{Sec_IV}

In this paper, we provide a possible explanation for the X-ray light curve of the FXT EP260321a during its SBO phase, which preceded the rise of the Type Ic-BL SN~2026gzf.
A bare WR star SBO is unlikely to account for EP260321a, as this scenario predicts a non-thermal X-ray flash lasting only tens of seconds, which is significantly shorter than the observed X-ray timescale. Our analysis further suggests that a compact-CSM interpretation is similarly disfavored. The tension between the compact-CSM model and the observations of EP260321a is subtle and necessitates careful examination. We assess the viability of this scenario from three perspectives:
\begin{itemize}
\item In the compact-CSM regime ($\tau \gg c/v_{\rm s}$), SBO occurs at the outer edge of the shell, and the rise time for the SBO luminosity is set by the light-travel time across the CSM ($\sim R_{\rm out}/c$). The observed rise time of $\sim 690$-second therefore implies a CSM radius of $\sim 300\,R_{\odot}$ and a blackbody temperature of $\sim 10^{5}\,{\rm K}$ ($\sim$10 eV). This is nearly an order of magnitude below the observed radiation temperature of $\sim 0.1$ keV, disfavoring the compact-CSM interpretation. 
\item The prior distributions adopted in our MCMC fitting already encompass the compact-CSM regime. For instance, the configuration with $R_{\rm in} = 200\,R_{\odot}$ and $R_{\rm wid} = 100\,R_{\odot}$ (corresponding to $R_{\rm out} = 300\,R_{\odot}$), can reach an optical depth as large as $\tau \approx 540 (\gg c / v_{\rm s})$, corresponding to a CSM mass of $5\times10^{-3}\,M_{\odot}$). Nevertheless, the posterior distribution presented in Figure~\ref{fig3_mcmc} places both the median model and its $68\%$ credible interval firmly within the extended-CSM regime, disfavoring the compact-CSM region.
\item Recent studies \citep{2026ApJ..1007L...9R, 2026arXiv260610014Y} have refined the compact-CSM model for estimating the radiation temperature by incorporating the analytical framework of \citet{2010ApJ...725..904N}, and argue that this scenario is consistent with the observations of EP260321a. However, a fundamental internal inconsistency remains: these studies calculate the X-ray luminosity using formulae valid in the compact-CSM regime ($\tau \gg c/v_{\rm s}$), yet their inferred CSM parameters fall in the optically thin regime ($\tau \leq c/v_{\rm s}$), where a well-defined SBO cannot occur. 
\end{itemize}

Alternatively, a SBO occurring within an extended CSM shell is consistent with both the observed early X-ray light curve and the temperature evolution of EP260321a. The proposed model comprises four stages: (1) ionization of the CSM by the SBO from the progenitor star prior to the collision of the SN ejecta with the CSM; (2) accumulation of radiative energy behind the forward shock prior to breakout; (3) diffusion of the SBO radiation through the unshocked CSM; and (4) a subsequent cooling phase as the shock sweeps through the entire CSM shell. We implemented the analytical framework of \citet{2010ApJ...725..904N} to compute the radiation temperature behind the forward shock, and the Monte Carlo method of \citet{2025ApJ...984...44H} to simulate photon diffusion through the unshocked CSM. The best-fit CSM parameters place strong constraints on the late-stage mass-loss history of the progenitor star. The CSM shell is located at a distance of approximately 150--600 $R_\odot$ from the progenitor. Assuming a typical WR wind velocity of 1000\,km\,s$^{-1}$, this distance corresponds to a mass-loss episode occurring roughly 1--5 days before the core collapse of the progenitor star.

The total ejected mass of approximately $5\times10^{-4}\ M_\odot$ indicates a mass-loss rate several orders of magnitude greater than that of a typical WR star's steady wind. This result adds to the growing body of evidence that stripped-envelope SN progenitors can undergo intense, episodic mass loss in the final days to weeks before explosion. The detection of the X-ray SBO of EP260321a, whose light curve is well characterized by photon diffusion through a radially extended CSM shell, provides a crucial observational probe of this late-stage progenitor evolution.

\begin{acknowledgments}

We thank X. Wu and L. Wang for helpful discussions on SN~2026gzf and EP260321a.
This work is supported by the National Natural Science Foundation of China (NSFC grants 12288102, 12403049, and 12033003). 
Maokai Hu acknowledges the support from China Postdoctoral Science Foundation (Certificate Number 2025T180872), the Postdoctoral Fellowship Program of CPSF under Grant Number GZB20240376, and the Shuimu Tsinghua Scholar Program 2024SM118. 
Y.\,Y's research is partially supported by the Tsinghua University Dushi Program.
X. Wang acknowledges the support from the Tencent Xplorer Prize.

\end{acknowledgments}

\bibliography{sample701}{}

@ARTICLE{2025NatAs...9.1073S,
       author = {{Sun}, H. and {Li}, W.-X. and {Liu}, L.-D. and {Gao}, H. and {Wang}, X.-F. and {Yuan}, W. and {Zhang}, B. and {Filippenko}, A.~V. and {Xu}, D. and {An}, T. and {Ai}, S. and {Brink}, T.~G. and {Liu}, Y. and {Liu}, Y.-Q. and {Wang}, C.-Y. and {Wu}, Q.-Y. and {Wu}, X.-F. and {Yang}, Y. and {Zhang}, B.-B. and {Zheng}, W.-K. and {Ahumada}, T. and {Dai}, Z.-G. and {Delaunay}, J. and {Elias-Rosa}, N. and {Benetti}, S. and {Fu}, S.-Y. and {Howell}, D.~A. and {Huang}, Y.-F. and {Kasliwal}, M.~M. and {Karambelkar}, V. and {Stein}, R. and {Lei}, W.-H. and {Lian}, T.-Y. and {Peng}, Z.-K. and {Frederiks}, D.~D. and {Ridnaia}, A.~V. and {Svinkin}, D.~S. and {Wang}, X.-Y. and {Wang}, A.-L. and {Wei}, D.-M. and {An}, J. and {Andrews}, M. and {Bai}, J.-M. and {Dai}, C.-Y. and {Ehgamberdiev}, S.~A. and {Fan}, Z. and {Farah}, J. and {Feng}, H.-C. and {Fynbo}, J.~P.~U. and {Guo}, W.-J. and {Guo}, Z. and {Hu}, M.-K. and {Hu}, J.-W. and {Jiang}, S.-Q. and {Jin}, J.-J. and {Li}, A. and {Li}, J.-D. and {Li}, R.-Z. and {Liang}, Y.-F. and {Ling}, Z.-X. and {Liu}, X. and {Mao}, J.-R. and {McCully}, C. and {Mirzaqulov}, D. and {Newsome}, M. and {Padilla Gonzalez}, E. and {Pan}, X. and {Terreran}, G. and {Tinyanont}, S. and {Wang}, B.-T. and {Wang}, L.-Z. and {Wen}, X.-D. and {Xiang}, D.-F. and {Xue}, S.-J. and {Yang}, J. and {Zhu}, Z.-P. and {Cai}, Z.-M. and {Castro-Tirado}, A.~J. and {Chen}, F.-S. and {Chen}, H.-L. and {Chen}, T.-X. and {Chen}, W. and {Chen}, Y.-H. and {Chen}, Y.-F. and {Chen}, Y. and {Cheng}, H.-Q. and {Cordier}, B. and {Cui}, C.-Z. and {Cui}, W.-W. and {Dai}, Y.-F. and {Fan}, D.-W. and {Feng}, H. and {Guan}, J. and {Han}, D.-W. and {Hou}, D.-J. and {Hu}, H.-B. and {Huang}, M.-H. and {Huo}, J. and {Jia}, S.-M. and {Jia}, Z.-Q. and {Jiang}, B.-W. and {Jin}, C.-C. and {Jin}, G. and {Kuulkers}, E. and {Li}, C.-K. and {Li}, D.-Y. and {Li}, J.-F. and {Li}, L.-H. and {Li}, M.-S. and {Li}, W. and {Li}, Z.-D. and {Liu}, C.-Z. and {Liu}, H.-Y. and {Liu}, H.-Q. and {Liu}, M.-J. and {Lu}, F.-J. and {Luo}, L.-D. and {Ma}, J. and {Mao}, X. and {Nandra}, K. and {O'Brien}, P. and {Pan}, H.-W. and {Rau}, A. and {Rea}, N. and {Sanders}, J. and {Song}, L.-M. and {Sun}, S.-L. and {Sun}, X.-J. and {Tan}, Y.-Y. and {Tang}, Q.-J. and {Tao}, Y.-H. and {Wang}, H. and {Wang}, J. and {Wang}, L. and {Wang}, W.-X. and {Wang}, Y.-L. and {Wang}, Y.-S. and {Xiong}, D.-R. and {Xu}, H.-T. and {Xu}, J.-J. and {Xu}, X.-P. and {Xu}, Y.-F. and {Xu}, Z. and {Xue}, C.-B. and {Xue}, Y.-L. and {Yan}, A.-L. and {Yang}, H.-N. and {Yang}, X.-T. and {Yang}, Y.-J. and {Zhang}, C. and {Zhang}, J. and {Zhang}, M. and {Zhang}, S.-N. and {Zhang}, W.-D. and {Zhang}, W.-J. and {Zhang}, Y.-H. and {Zhang}, Z. and {Zhang}, Z. and {Zhang}, Z.-L. and {Zhao}, D.-H. and {Zhao}, H.-S. and {Zhao}, X.-F. and {Zhao}, Z.-J. and {Zhou}, Y.-L. and {Zhu}, Y.-X. and {Zhu}, Z.-C. and {Zou}, H.},
        title = "{A fast X-ray transient from a weak relativistic jet associated with a type Ic-BL supernova}",
      journal = {Nature Astronomy},
         year = 2025,
        month = jul,
       volume = {9},
        pages = {1073-1085},
          doi = {10.1038/s41550-025-02571-1},
archivePrefix = {arXiv},
       eprint = {2410.02315},
 primaryClass = {astro-ph.HE},
       adsurl = {https://ui.adsabs.harvard.edu/abs/2025NatAs...9.1073S}
}

@ARTICLE{2025ApJ...978L..27H,
       author = {{Hu}, Maokai and {Ao}, Yiping and {Yang}, Yi and {Hu}, Lei and {Li}, Fulin and {Wang}, Lifan and {Wang}, Xiaofeng},
        title = "{Early-time Millimeter Observations of the Nearby Type II SN 2024ggi}",
      journal = {\apjl},
         year = 2025,
        month = jan,
       volume = {978},
       number = {2},
          eid = {L27},
        pages = {L27},
          doi = {10.3847/2041-8213/ada1cd},
archivePrefix = {arXiv},
       eprint = {2412.11389},
 primaryClass = {astro-ph.SR},
       adsurl = {https://ui.adsabs.harvard.edu/abs/2025ApJ...978L..27H}
}

@ARTICLE{2025ApJ...984...44H,
       author = {{Hu}, Maokai and {Wang}, Lifan and {Wang}, Xiaofeng},
        title = "{A Shock Crashing into Confined Dense Circumstellar Matter Brightens the Nascent SN 2023ixf}",
      journal = {\apj},
         year = 2025,
        month = may,
       volume = {984},
       number = {1},
          eid = {44},
        pages = {44},
          doi = {10.3847/1538-4357/adc802},
archivePrefix = {arXiv},
       eprint = {2411.06351},
 primaryClass = {astro-ph.HE},
       adsurl = {https://ui.adsabs.harvard.edu/abs/2025ApJ...984...44H}
}

@ARTICLE{2023MNRAS.525..246H,
       author = {{Hu}, Maokai and {Wang}, Lifan and {Wang}, Xiaofeng and {Wang}, Lingzhi},
        title = "{Possible circumstellar interaction origin of the early excess emission in thermonuclear supernovae}",
      journal = {\mnras},
         year = 2023,
        month = oct,
       volume = {525},
       number = {1},
        pages = {246-255},
          doi = {10.1093/mnras/stad2340},
archivePrefix = {arXiv},
       eprint = {2307.02186},
 primaryClass = {astro-ph.HE},
       adsurl = {https://ui.adsabs.harvard.edu/abs/2023MNRAS.525..246H}
}

@ARTICLE{2008Natur.453..469S,
       author = {{Soderberg}, A.~M. and {Berger}, E. and {Page}, K.~L. and {Schady}, P. and {Parrent}, J. and {Pooley}, D. and {Wang}, X.-Y. and {Ofek}, E.~O. and {Cucchiara}, A. and {Rau}, A. and {Waxman}, E. and {Simon}, J.~D. and {Bock}, D.~C.-J. and {Milne}, P.~A. and {Page}, M.~J. and {Barentine}, J.~C. and {Barthelmy}, S.~D. and {Beardmore}, A.~P. and {Bietenholz}, M.~F. and {Brown}, P. and {Burrows}, A. and {Burrows}, D.~N. and {Byrngelson}, G. and {Cenko}, S.~B. and {Chandra}, P. and {Cummings}, J.~R. and {Fox}, D.~B. and {Gal-Yam}, A. and {Gehrels}, N. and {Immler}, S. and {Kasliwal}, M. and {Kong}, A.~K.~H. and {Krimm}, H.~A. and {Kulkarni}, S.~R. and {Maccarone}, T.~J. and {M{\'e}sz{\'a}ros}, P. and {Nakar}, E. and {O'Brien}, P.~T. and {Overzier}, R.~A. and {de Pasquale}, M. and {Racusin}, J. and {Rea}, N. and {York}, D.~G.},
        title = "{An extremely luminous X-ray outburst at the birth of a supernova}",
      journal = {\nat},
         year = 2008,
        month = may,
       volume = {453},
       number = {7194},
        pages = {469-474},
          doi = {10.1038/nature06997},
archivePrefix = {arXiv},
       eprint = {0802.1712},
 primaryClass = {astro-ph},
       adsurl = {https://ui.adsabs.harvard.edu/abs/2008Natur.453..469S}
}

@ARTICLE{2010ApJ...725..904N,
       author = {{Nakar}, Ehud and {Sari}, Re'em},
        title = "{Early Supernovae Light Curves Following the Shock Breakout}",
      journal = {\apj},
         year = 2010,
        month = dec,
       volume = {725},
       number = {1},
        pages = {904-921},
          doi = {10.1088/0004-637X/725/1/904},
archivePrefix = {arXiv},
       eprint = {1004.2496},
 primaryClass = {astro-ph.HE},
       adsurl = {https://ui.adsabs.harvard.edu/abs/2010ApJ...725..904N}
}

@ARTICLE{2011ApJ...729L...6C,
       author = {{Chevalier}, Roger A. and {Irwin}, Christopher M.},
        title = "{Shock Breakout in Dense Mass Loss: Luminous Supernovae}",
      journal = {\apjl},
         year = 2011,
        month = mar,
       volume = {729},
       number = {1},
          eid = {L6},
        pages = {L6},
          doi = {10.1088/2041-8205/729/1/L6},
archivePrefix = {arXiv},
       eprint = {1101.1111},
 primaryClass = {astro-ph.HE},
       adsurl = {https://ui.adsabs.harvard.edu/abs/2011ApJ...729L...6C}
}

@ARTICLE{2013PASP..125..306F,
       author = {{Foreman-Mackey}, Daniel and {Hogg}, David W. and {Lang}, Dustin and {Goodman}, Jonathan},
        title = "{emcee: The MCMC Hammer}",
      journal = {\pasp},
         year = 2013,
        month = mar,
       volume = {125},
       number = {925},
        pages = {306},
          doi = {10.1086/670067},
archivePrefix = {arXiv},
       eprint = {1202.3665},
 primaryClass = {astro-ph.IM},
       adsurl = {https://ui.adsabs.harvard.edu/abs/2013PASP..125..306F}
}

@ARTICLE{2022ApJ...933..238M,
       author = {{Margalit}, Ben},
        title = "{Analytic Light Curves of Dense CSM Shock Breakout and Cooling}",
      journal = {\apj},
         year = 2022,
        month = jul,
       volume = {933},
       number = {2},
          eid = {238},
        pages = {238},
          doi = {10.3847/1538-4357/ac771a},
archivePrefix = {arXiv},
       eprint = {2107.04048},
 primaryClass = {astro-ph.HE},
       adsurl = {https://ui.adsabs.harvard.edu/abs/2022ApJ...933..238M}
}

@ARTICLE{2026arXiv260610014Y,
       author = {{Yuan}, Weimin and {Huang}, Qiu-Ju and {Zhu}, Jin-Ping and {Yu}, Yun-Wei and {Xu}, Dong and {Zhang}, Chen and {Li}, Zhuo and {Liu}, Yuan and {An}, Tao and {Gianfagna}, Giulia and {Zheng}, Weikang and {Du}, Guowang and {Liu}, Xing and {Jiang}, Ji-An and {Fynbo}, Johan P.~U. and {Pozanenko}, Alexei S. and {Jin}, Junjie and {Yang}, Yi and {Deng}, Jinsong and {Sun}, Hui and {Wu}, Guang-Lei and {Zhang}, Yu-Hao and {Wang}, Bao and {Wang}, Yu and {Wang}, Xiangyu and {Zhang}, Bin-Bin and {Chen}, Yong and {Zhang}, Yonghe and {Wang}, Bo and {Wang}, Xiaofeng and {Wu}, Xuefeng and {Dai}, Zigao and {An}, Jie and {Anupama}, G.~C. and {Balasubramanian}, Arvind and {Bao}, Congying and {Beri}, Aru and {Bhalerao}, Varun and {Brink}, Thomas G. and {Bruni}, Gabriele and {Cai}, Minxuan and {Cai}, Zhiming and {Chanchaiworawit}, Krittapas and {Chen}, Yehai and {Cheng}, Huaqing and {Cordier}, Bertrand and {Cui}, Chenzhou and {Cui}, Weiwei and {Dai}, Cuiyuan and {Eappachen}, D. and {Eselevich}, M.~V. and {Fan}, Xiao and {Fan}, Zhou and {Fang}, Yuan and {Feng}, Hua and {Filippenko}, Alexei V. and {Fu}, Shaoyu and {Gao}, He and {Geng}, Jinjun and {Goranskij}, Vitaly and {Guan}, Ju and {Han}, Dawei and {Hao}, Jinxin and {He}, Linbo and {He}, Min and {Hu}, Jingwei and {Huang}, Maohai and {Jia}, Shumei and {Jia}, Ziqing and {Jiang}, Shuaiqing and {Jin}, Chichuan and {Jin}, Ge and {Jonker}, Peter and {Klunko}, E.~V. and {Kong}, Albert K.~H. and {Li}, Chengkui and {Li}, Dongyue and {Li}, Rui-Zhi and {Li}, Wenxiong and {Liang}, Run-Duo and {Ling}, Zhixing and {Liu}, Congzhan and {Liu}, Huaqiu and {Liu}, Liangduan and {Liu}, Xiangkun and {Liu}, Xiaowei and {Liu}, Yuanqi and {Liu}, Zhengwei and {Lu}, Fangjun and {Mao}, Jirong and {Mao}, Xuan and {Moskvitin}, A.~S. and {Mu}, Haiyang and {Nandra}, Kirpal and {Ness}, Jan-Uwe and {Ni}, Kangrui and {Noysena}, Kanthanakorn and {O'Brien}, Paul and {Pan}, Haiwu and {Pan}, Yu and {Pankov}, N.~S. and {Piro}, Luigi and {Quirola-Vasquez}, J. and {Rau}, Arne and {Rea}, Nanda and {Sahu}, D.~K. and {Pawan Saikia}, Aditya and {Sanders}, Jeremy and {Song}, Liming and {Spiridonova}, Olga and {Sun}, Ning-Chen and {Sun}, Shengli and {Sun}, Xiaojin and {Tan}, Yuyin and {Thakur}, Aishwarya Linesh and {Tinyanont}, Samaporn and {Vlasyuk}, Valery and {Volnova}, A.~V. and {Wang}, Ailing and {Wu}, Hong and {Wu}, Qianrui and {Xu}, Haitao and {Xu}, Zelin and {Xue}, Changbin and {Yin}, Yi-Han Iris and {Zaznobin}, I.~A. and {Zhang}, Jia-Sen and {Zhang}, Shuang-Nan and {Zhang}, Songbo and {Zhang}, Yu and {Zhu}, Zipei and {Zou}, Zecheng and {Zhang}, Bing},
        title = "{X-rays breaking out of pre-explosion ejecta mark a supernova's first light}",
      journal = {arXiv e-prints},
         year = 2026,
        month = jun,
          eid = {arXiv:2606.10014},
        pages = {arXiv:2606.10014},
          doi = {10.48550/arXiv.2606.10014},
archivePrefix = {arXiv},
       eprint = {2606.10014},
 primaryClass = {astro-ph.HE},
       adsurl = {https://ui.adsabs.harvard.edu/abs/2026arXiv260610014Y}
}

@ARTICLE{2026ApJ..1006L..13O,
       author = {{O'Connor}, Brendan and {Hall}, Xander J. and {Busmann}, Malte and {Gruen}, Daniel and {Floris}, Alberto and {Cabrera}, Tom{\'a}s and {Zhu}, Ziyuan and {Palmese}, Antonella and {Green}, Dylan and {Banovetz}, John and {Gassert}, Julius and {Fryer}, Christopher L. and {Ricci}, Roberto and {Troja}, Eleonora and {Shivaprasad}, Surya and {Zeimann}, Gregory R. and {Amsellem}, Ariel J. and {Bailey}, Stephen and {BenZvi}, Segev and {Dichiara}, Simone and {van Eerten}, Hendrik and {Hare}, Jeremy and {Hu}, Lei and {Irwin}, Christopher M. and {Kunnumkai}, Keerthi and {Malanchev}, Konstantin and {Maleki}, Mitra and {Moss}, Michael J. and {Myers}, Adam D. and {Pasham}, Dheeraj and {Ries}, Christoph and {Ryan}, Geoffrey and {Schlegel}, David and {Schmidt}, Michael and {Wilke}, Silona and {Yang}, Yu-Han},
        title = "{EP260321a/SN 2026gzf: The Faintest Shock Breakout Associated with a Broad-lined Supernova}",
      journal = {\apjl},
         year = 2026,
        month = jul,
       volume = {1006},
       number = {1},
          eid = {L13},
        pages = {L13},
          doi = {10.3847/2041-8213/ae84ba},
archivePrefix = {arXiv},
       eprint = {2606.09992},
 primaryClass = {astro-ph.HE},
       adsurl = {https://ui.adsabs.harvard.edu/abs/2026ApJ..1006L..13O}
}

@ARTICLE{2026ApJ..1008L...3W,
       author = {{Wen}, Xudong and {Yang}, Yi and {Lu}, Jing and {Wang}, Lifan and {Wheeler}, J. Craig and {Pursiainen}, Miika and {Huang}, QiuJu and {Wang}, Bao and {Wu}, Xuefeng and {Filippenko}, Alexei V. and {Kerzendorf}, Wolfgang E. and {Leloudas}, Giorgos and {Schulze}, Steve and {Patat}, Ferdinando},
        title = "{Pinning Down the Geometry of the Type Ic Broad-line Supernova 2026gzf}",
      journal = {\apjl},
         year = 2026,
        month = sep,
       volume = {1008},
       number = {1},
          eid = {L3},
        pages = {L3},
          doi = {10.3847/2041-8213/ae990c},
archivePrefix = {arXiv},
       eprint = {2606.18881},
 primaryClass = {astro-ph.HE},
       adsurl = {https://ui.adsabs.harvard.edu/abs/2026ApJ..1008L...3W}
}

@ARTICLE{2026arXiv260610002M,
       author = {{Martin-Carrillo}, Antonio and {Th{\"o}ne}, Christina C. and {Leung}, James K. and {Corcoran}, Gregory and {de Ugarte Postigo}, Antonio and {Jonker}, Peter G. and {Izzo}, Luca and {Levan}, Andrew J. and {Gompertz}, Benjamin P. and {Basa}, St{\'e}phane and {Sarin}, Nikhil and {Quirola-V{\'a}squez}, Jonathan and {Eyles-Ferris}, Rob A.~J. and {Brivio}, Riccardo and {Watson}, Alan M. and {Cotter}, Laura and {Chac{\'o}n}, Jennifer Alexandra and {Rossi}, Andrea and {Melandri}, Andrea and {Kumnurdmanee}, Piramon and {Tanvir}, Nial R. and {Gupta}, Anshika and {Bauer}, Franz E. and {Ducoin}, Jean-Gr{\'e}goire and {Reguitti}, Andrea and {Misra}, Kuntal and {Xu}, Dong and {Vergani}, Susanna D. and {Fong}, Wen-fai and {Ackley}, Kendall and {Aguilar-Ruiz}, Edilberto and {Akl}, Dalya and {Aloy}, Miguel {\'A}ngel and {An}, Jie and {Angulo-Valdez}, Camila and {Antier}, Sarah and {Atteia}, Jean-Luc and {Becerra}, Rosa L. and {Breton}, Rene P. and {Butler}, Nathaniel R. and {Campana}, Sergio and {Carotenuto}, Francesco and {Casares Vel{\'a}zquez}, Jorge and {Chrimes}, Ashley A. and {D'Elia}, Valerio and {van Dalen}, Joyce N.~D. and {De Colle}, Fabio and {De Pasquale}, Massimiliano and {Dhillon}, Vik S. and {Dornic}, Damien and {Dyer}, Martin J. and {Ferro}, Matteo and {Fraser}, Morgan and {Fruchter}, Andrew S. and {Fortin}, Francis and {Galloway}, Duncan K. and {Garc{\'\i}a-Garc{\'\i}a}, Leonardo and {Geier}, Stefan and {Gill}, Ramandeep and {Globus}, No{\'e}mie and {Gualandi}, Roberto and {Guelfand}, Marion and {Guidolin}, Francesco and {Hartmann}, Dieter H. and {van Hoof}, Agnes P.~C. and {Jakobsson}, Pall and {Janghel}, Divyanshu and {Killestein}, Tom L. and {Klose}, Sylvio and {Kobayashi}, Shiho and {Kotak}, Rubina and {Kumar}, Amit and {Kuwata}, Asuka and {Laskar}, Tanmoy and {Lee}, William H. and {Lincetto}, Massimiliano and {Lombardi}, Gianluca and {L{\'o}pez-C{\'a}mara}, Diego and {Lyman}, Joseph D. and {Maiorano}, Elisabetta and {Maeda}, Keiichi and {Mandarakas}, Nikos and {Magnani}, Francesco and {Mao}, Jirong and {Moreno M{\'e}ndez}, Enrique and {Mar{\'\i}a Nicuesa Guelbenzu}, Ana and {Noysena}, Kanthanakorn and {Nuttall}, Laura K. and {O'Brien}, Paul T. and {O'Neill}, David and {Ochner}, Paolo and {Pereyra}, Margarita and {Pugliese}, Giovanna and {Ramsay}, Gavin and {Rhodes}, Lauren and {Saccardi}, Andrea and {Salvaterra}, Ruben and {S{\'a}nchez {\'A}lvarez}, Fredd and {Schneider}, Benjamin and {Schulze}, Steve and {Starling}, Rhaana L.~C. and {Steeghs}, Danny and {Ulaczyk}, Kzrysztof and {Ventura}, Chiara and {Zafar}, Tayyaba and {Zhu}, Zi-Pei},
        title = "{Failed jet breakout in the metal-poor broad-lined type Ic supernova 2026gzf}",
      journal = {arXiv e-prints},
         year = 2026,
        month = jun,
          eid = {arXiv:2606.10002},
        pages = {arXiv:2606.10002},
          doi = {10.48550/arXiv.2606.10002},
archivePrefix = {arXiv},
       eprint = {2606.10002},
 primaryClass = {astro-ph.HE},
       adsurl = {https://ui.adsabs.harvard.edu/abs/2026arXiv260610002M}
}

@ARTICLE{2021ApJ...910..128H,
       author = {{Haynie}, Annastasia and {Piro}, Anthony L.},
        title = "{Shock Breakout in Dense Circumstellar Material with Application to PS1-13arp}",
      journal = {\apj},
         year = 2021,
        month = apr,
       volume = {910},
       number = {2},
          eid = {128},
        pages = {128},
          doi = {10.3847/1538-4357/abe938},
archivePrefix = {arXiv},
       eprint = {2011.01937},
 primaryClass = {astro-ph.HE},
       adsurl = {https://ui.adsabs.harvard.edu/abs/2021ApJ...910..128H}
}

@ARTICLE{2022ApJ...928..122M,
       author = {{Margalit}, Ben and {Quataert}, Eliot and {Ho}, Anna Y.~Q.},
        title = "{Optical to X-Ray Signatures of Dense Circumstellar Interaction in Core-collapse Supernovae}",
      journal = {\apj},
         year = 2022,
        month = apr,
       volume = {928},
       number = {2},
          eid = {122},
        pages = {122},
          doi = {10.3847/1538-4357/ac53b0},
archivePrefix = {arXiv},
       eprint = {2109.09746},
 primaryClass = {astro-ph.HE},
       adsurl = {https://ui.adsabs.harvard.edu/abs/2022ApJ...928..122M}
}

@ARTICLE{2026arXiv260610009C,
       author = {{Chen}, Ting-Wan and {Aryan}, Amar and {Yang}, Sheng and {Smartt}, Stephen J. and {Moriya}, Takashi J. and {Brennan}, Se{\'a}n J. and {Stritzinger}, Maximilian D. and {Martin}, Bailey and {Nicholl}, Matt and {Kong}, Albert K.~H. and {Gillanders}, James H. and {Dutta}, Anirban and {Schmidt}, Brian P. and {Cheng}, Yu-Chi and {Huber}, Mark E. and {Lai}, Cheng-Han and {Lee}, Chien-Hsiu and {Lee}, Yu-Hsing and {Ngeow}, Chow-Choong and {Smith}, Ken W. and {Ashall}, Christopher and {Auchettl}, Katie and {Burns}, Chris R. and {Chambers}, Kenneth C. and {Chen}, Zhi-Yue and {de Boer}, Thomas and {Hsiao}, Eric Y. and {Ngo Thanh Ho}, Khoa and {Hoogendam}, Willem B. and {Jones}, David O. and {Kankare}, Erkki and {Killestein}, Tom L. and {Kuncarayakti}, Hanindyo and {Lee}, Meng-Han and {Li}, Chuan-Jui and {Lin}, Chien-Cheng and {Lidman}, Christopher and {Lowe}, Thomas B. and {Magnier}, Eugene A. and {Medler}, Kyle and {M{\"o}ller}, Anais and {Moore}, Thomas and {Morrell}, Nidia and {Paek}, Gregory S.~H. and {Pfeffer}, Cameron M. and {Qiang}, Da-Chun and {Rauf}, Liana and {Reynolds}, Thomas M. and {Sankar. K}, Aiswarya and {Srivastav}, Shubham and {Tweddle}, Jack and {Wainscoat}, Richard and {Wang}, Ze-Ning and {Xiao}, Huangfei and {Zhu}, Zonghong},
        title = "{Decadal pre-explosion activity and circumstellar interaction in a supernova}",
      journal = {arXiv e-prints},
         year = 2026,
        month = jun,
          eid = {arXiv:2606.10009},
        pages = {arXiv:2606.10009},
          doi = {10.48550/arXiv.2606.10009},
archivePrefix = {arXiv},
       eprint = {2606.10009},
 primaryClass = {astro-ph.HE},
       adsurl = {https://ui.adsabs.harvard.edu/abs/2026arXiv260610009C}
}

@ARTICLE{2026ApJ..1007L...9R,
       author = {{Rastinejad}, Jillian C. and {Srinivasaragavan}, Gokul and {Sarin}, Nikhil and {O'Dwyer}, Tanner and {Cenko}, S. Bradley and {Leung}, James K. and {Nugent}, Anya E. and {Perley}, Daniel A. and {Schroeder}, Genevieve and {Anand}, Shreya and {Ahumada}, Tom{\'a}s and {Andreoni}, Igor and {Bochenek}, Aleksandra and {Corsi}, Alessandra and {Fremling}, Christoffer and {Ho}, Anna Y.~Q. and {Kasliwal}, Mansi M. and {Mo}, Geoffrey and {Salgundi}, Anirudh and {Sippy}, Kendall I. and {Sollerman}, J. and {Bellm}, Eric C. and {Chen}, Tracy X. and {Coughlin}, Michael W. and {Davis}, Michael C. and {De Colle}, Fabio and {Frostig}, Danielle and {Fryer}, Christopher L. and {Graham}, Michael J. and {Hall}, Xander J. and {Hinds}, K.-R. and {Izzo}, Luca and {Jacobson-Gal{\'a}n}, Wynn and {Lourie}, Nathan P. and {Maeda}, Keiichi and {Purdum}, Josiah and {Rusholme}, Ben and {Singh}, Avinash and {Stein}, Robert},
        title = "{A Multiwavelength View of the First Type Ic-BL Supernova with an Einstein Probe X-Ray Shock Breakout}",
      journal = {\apjl},
         year = 2026,
        month = aug,
       volume = {1007},
       number = {1},
          eid = {L9},
        pages = {L9},
          doi = {10.3847/2041-8213/ae8a4b},
       adsurl = {https://ui.adsabs.harvard.edu/abs/2026ApJ..1007L...9R}
}

@ARTICLE{2015arXiv150607735Y,
       author = {{Yuan}, Weimin and {Zhang}, C. and {Feng}, H. and {Zhang}, S.~N. and {Ling}, Z.~X. and {Zhao}, D. and {Deng}, J. and {Qiu}, Y. and {Osborne}, J.~P. and {O'Brien}, P. and {Willingale}, R. and {Lapington}, J. and {Fraser}, G.~W. and {the Einstein Probe team}},
        title = "{Einstein Probe - a small mission to monitor and explore the dynamic X-ray Universe}",
      journal = {arXiv e-prints},
         year = 2015,
        month = jun,
          eid = {arXiv:1506.07735},
        pages = {arXiv:1506.07735},
          doi = {10.48550/arXiv.1506.07735},
archivePrefix = {arXiv},
       eprint = {1506.07735},
 primaryClass = {astro-ph.HE},
       adsurl = {https://ui.adsabs.harvard.edu/abs/2015arXiv150607735Y}
}

@INCOLLECTION{2022hxga.book...86Y,
       author = {{Yuan}, Weimin and {Zhang}, Chen and {Chen}, Yong and {Ling}, Zhixing},
        title = "{The Einstein Probe Mission}",
    booktitle = {Handbook of X-ray and Gamma-ray Astrophysics},
         year = 2022,
       editor = {{Bambi}, Cosimo and {Sangangelo}, Andrea},
          eid = {86},
        pages = {86},
          doi = {10.1007/978-981-16-4544-0_151-1},
       adsurl = {https://ui.adsabs.harvard.edu/abs/2022hxga.book...86Y}
}

@ARTICLE{2025ApJ...982L..47V,
       author = {{van Dalen}, Joyce N.~D. and {Levan}, Andrew J. and {Jonker}, Peter G. and {Malesani}, Daniele Bj{\o}rn and {Izzo}, Luca and {Sarin}, Nikhil and {Quirola-V{\'a}squez}, Jonathan and {Mata S{\'a}nchez}, Daniel and {de Ugarte Postigo}, Antonio and {van Hoof}, Agnes P.~C. and {Torres}, Manuel A.~P. and {Schulze}, Steve and {Littlefair}, Stuart P. and {Chrimes}, Ashley and {Ravasio}, Maria E. and {Bauer}, Franz E. and {Martin-Carrillo}, Antonio and {Fraser}, Morgan and {van der Horst}, Alexander J. and {Jakobsson}, Pall and {O'Brien}, Paul and {De Pasquale}, Massimiliano and {Pugliese}, Giovanna and {Sollerman}, Jesper and {Tanvir}, Nial R. and {Zafar}, Tayyaba and {Anderson}, Joseph P. and {Galbany}, Llu{\'\i}s and {Gal-Yam}, Avishay and {Gromadzki}, Mariusz and {M{\"u}ller-Bravo}, Tom{\'a}s E. and {Ragosta}, Fabio and {Terwel}, Jacco H.},
        title = "{The Einstein Probe Transient EP240414a: Linking Fast X-Ray Transients, Gamma-Ray Bursts, and Luminous Fast Blue Optical Transients}",
      journal = {\apjl},
         year = 2025,
        month = apr,
       volume = {982},
       number = {2},
          eid = {L47},
        pages = {L47},
          doi = {10.3847/2041-8213/adbc7e},
archivePrefix = {arXiv},
       eprint = {2409.19056},
 primaryClass = {astro-ph.HE},
       adsurl = {https://ui.adsabs.harvard.edu/abs/2025ApJ...982L..47V}
}

@ARTICLE{2025ApJ...988L..14E,
       author = {{Eyles-Ferris}, Rob A.~J. and {Jonker}, Peter G. and {Levan}, Andrew J. and {Malesani}, Daniele Bj{\o}rn and {Sarin}, Nikhil and {Fryer}, Christopher L. and {Rastinejad}, Jillian C. and {Burns}, Eric and {Tanvir}, Nial R. and {O'Brien}, Paul T. and {Fong}, Wen-fai and {Mandel}, Ilya and {Gompertz}, Benjamin P. and {Kilpatrick}, Charles D. and {Bloemen}, Steven and {Bright}, Joe S. and {Carotenuto}, Francesco and {Corcoran}, Gregory and {Cotter}, Laura and {Groot}, Paul J. and {Izzo}, Luca and {Laskar}, Tanmoy and {Martin-Carrillo}, Antonio and {Palmerio}, Jesse and {Ravasio}, Maria E. and {van Roestel}, Jan and {Saccardi}, Andrea and {Starling}, Rhaana L.~C. and {Thakur}, Aishwarya Linesh and {Vergani}, Susanna D. and {Vreeswijk}, Paul M. and {Bauer}, Franz E. and {Campana}, Sergio and {Chac{\'o}n}, Jennifer A. and {Chrimes}, Ashley A. and {Covino}, Stefano and {van Dalen}, Joyce N.~D. and {D'Elia}, Valerio and {De Pasquale}, Massimiliano and {Habeeb}, Nusrin and {Hartmann}, Dieter H. and {van Hoof}, Agnes P.~C. and {Jakobsson}, P{\'a}ll and {Julakanti}, Yashaswi and {Leloudas}, Giorgos and {Mata S{\'a}nchez}, Daniel and {Nixon}, Christopher J. and {Pieterse}, Dani{\"e}lle L.~A. and {Pugliese}, Giovanna and {Quirola-V{\'a}squez}, Jonathan and {Rayson}, Ben C. and {Salvaterra}, Ruben and {Schneider}, Ben and {Torres}, Manuel A.~P. and {Zafar}, Tayyaba},
        title = "{The Kangaroo's First Hop: The Early Fast Cooling Phase of EP250108a/SN 2025kg}",
      journal = {\apjl},
         year = 2025,
        month = jul,
       volume = {988},
       number = {1},
          eid = {L14},
        pages = {L14},
          doi = {10.3847/2041-8213/ade1d9},
archivePrefix = {arXiv},
       eprint = {2504.08886},
 primaryClass = {astro-ph.HE},
       adsurl = {https://ui.adsabs.harvard.edu/abs/2025ApJ...988L..14E}
}

@ARTICLE{2025ApJ...986L...4H,
       author = {{Hamidani}, Hamid and {Sato}, Yuri and {Kashiyama}, Kazumi and {Tanaka}, Masaomi and {Ioka}, Kunihito and {Kimura}, Shigeo S.},
        title = "{EP240414a: A Gamma-Ray Burst Jet Weakened by an Extended Circumstellar Material}",
      journal = {\apjl},
         year = 2025,
        month = jun,
       volume = {986},
       number = {1},
          eid = {L4},
        pages = {L4},
          doi = {10.3847/2041-8213/add99d},
archivePrefix = {arXiv},
       eprint = {2503.16243},
 primaryClass = {astro-ph.HE},
       adsurl = {https://ui.adsabs.harvard.edu/abs/2025ApJ...986L...4H}
}

@ARTICLE{2025ApJ...988L..13R,
       author = {{Rastinejad}, Jillian C. and {Levan}, Andrew J. and {Jonker}, Peter G. and {Kilpatrick}, Charles D. and {Fryer}, Christopher L. and {Sarin}, Nikhil and {Gompertz}, Benjamin P. and {Liu}, Chang and {Eyles-Ferris}, Rob A.~J. and {Fong}, Wen-fai and {Burns}, Eric and {Gillanders}, James H. and {Mandel}, Ilya and {Malesani}, Daniele Bj{\o}rn and {O'Brien}, Paul T. and {Tanvir}, Nial R. and {Ackley}, Kendall and {Aryan}, Amar and {Bauer}, Franz E. and {Bloemen}, Steven and {de Boer}, Thomas and {Bom}, Cl{\'e}cio R. and {Chac{\'o}n}, Jennifer A. and {Chambers}, Ken and {Chen}, Ting-Wan and {Chrimes}, Ashley A. and {van Dalen}, Joyce N.~D. and {D'Elia}, Valerio and {De Pasquale}, Massimiliano and {Fulton}, Michael D. and {Groot}, Paul J. and {Gupta}, Rahul and {Hartmann}, Dieter H. and {van Hoof}, Agnes P.~C. and {Huber}, Mark E. and {Izzo}, Luca and {Jacobson-Galan}, Wynn and {Jakobsson}, P{\'a}ll and {Kong}, Albert and {Laskar}, Tanmoy and {Lowe}, Thomas B. and {Magnier}, Eugene A. and {Maiorano}, Elisabetta and {Martin-Carrillo}, Antonio and {Mas-Ribas}, Lluis and {Mata S{\'a}nchez}, Daniel and {Nicholl}, Matt and {Nixon}, Christopher J. and {Oates}, Samantha R. and {Paek}, Gregory and {Palmerio}, Jesse and {Paris}, Diego and {Pieterse}, Dani{\"e}lle L.~A. and {Pugliese}, Giovanna and {Quirola Vasquez}, Jonathan A. and {van Roestel}, Jan and {Rossi}, Andrea and {Rouco Escorial}, Alicia and {Salvaterra}, Ruben and {Schneider}, Benjamin and {Smartt}, Stephen J. and {Smith}, Ken and {Smith}, Ian A. and {Srivastav}, Shubham and {Torres}, Manuel A.~P. and {Ventura}, Chiara and {Vreeswijk}, Paul and {Wainscoat}, Richard and {Yang}, Yi-Jung and {Yang}, Sheng},
        title = "{EP 250108a/SN 2025kg: Observations of the Most Nearby Broad-line Type Ic Supernova Following an Einstein Probe Fast X-Ray Transient}",
      journal = {\apjl},
         year = 2025,
        month = jul,
       volume = {988},
       number = {1},
          eid = {L13},
        pages = {L13},
          doi = {10.3847/2041-8213/ade7f9},
archivePrefix = {arXiv},
       eprint = {2504.08889},
 primaryClass = {astro-ph.HE},
       adsurl = {https://ui.adsabs.harvard.edu/abs/2025ApJ...988L..13R}
}

@ARTICLE{2026ApJ...996L..29V,
       author = {{van Hoof}, Agnes P.~C. and {Levan}, Andrew J. and {Jonker}, Peter G. and {Fraser}, Morgan and {Izzo}, Luca and {Fruchter}, Andrew S. and {van Dalen}, Joyce N.~D. and {Tanvir}, Nial R. and {Hjorth}, Jens and {Martin-Carrillo}, Antonio and {Sarin}, Nikhil and {Cotter}, Laura C. and {Quirola Vasquez}, Jonathan A. and {Ravasio}, Maria E. and {S{\'a}nchez-Sierras}, Javi and {Torres}, Manuel A.~P. and {Jiang}, Shuai-Qing and {Xu}, Dong},
        title = "{JWST and HST Observations of the Host Galaxy and Supernova SN 2024aihh in EP240801a at z = 1.67}",
      journal = {\apjl},
         year = 2026,
        month = jan,
       volume = {996},
       number = {2},
          eid = {L29},
        pages = {L29},
          doi = {10.3847/2041-8213/ae28db},
archivePrefix = {arXiv},
       eprint = {2512.11572},
 primaryClass = {astro-ph.HE},
       adsurl = {https://ui.adsabs.harvard.edu/abs/2026ApJ...996L..29V}
}

@ARTICLE{1978ApJ...223L.109K,
       author = {{Klein}, R.~I. and {Chevalier}, R.~A.},
        title = "{X-ray bursts from type II supernovae.}",
      journal = {\apjl},
         year = 1978,
        month = aug,
       volume = {223},
        pages = {L109-L112},
          doi = {10.1086/182740},
       adsurl = {https://ui.adsabs.harvard.edu/abs/1978ApJ...223L.109K}
}

@ARTICLE{1992ApJ...393..742E,
       author = {{Ensman}, Lisa and {Burrows}, Adam},
        title = "{Shock Breakout in SN 1987A}",
      journal = {\apj},
         year = 1992,
        month = jul,
       volume = {393},
        pages = {742},
          doi = {10.1086/171542},
       adsurl = {https://ui.adsabs.harvard.edu/abs/1992ApJ...393..742E}
}

@ARTICLE{1999ApJ...510..379M,
       author = {{Matzner}, Christopher D. and {McKee}, Christopher F.},
        title = "{The Expulsion of Stellar Envelopes in Core-Collapse Supernovae}",
      journal = {\apj},
         year = 1999,
        month = jan,
       volume = {510},
       number = {1},
        pages = {379-403},
          doi = {10.1086/306571},
archivePrefix = {arXiv},
       eprint = {astro-ph/9807046},
 primaryClass = {astro-ph},
       adsurl = {https://ui.adsabs.harvard.edu/abs/1999ApJ...510..379M}
}

@ARTICLE{2012ApJ...747..147K,
       author = {{Katz}, Boaz and {Sapir}, Nir and {Waxman}, Eli},
        title = "{Non-relativistic Radiation Mediated Shock Breakouts. II. Bolometric Properties of Supernova Shock Breakout}",
      journal = {\apj},
         year = 2012,
        month = mar,
       volume = {747},
       number = {2},
          eid = {147},
        pages = {147},
          doi = {10.1088/0004-637X/747/2/147},
archivePrefix = {arXiv},
       eprint = {1103.5276},
 primaryClass = {astro-ph.HE},
       adsurl = {https://ui.adsabs.harvard.edu/abs/2012ApJ...747..147K}
}

@ARTICLE{2020MNRAS.494.3927K,
       author = {{Kozyreva}, Alexandra and {Nakar}, Ehud and {Waldman}, Roni and {Blinnikov}, Sergei and {Baklanov}, Petr},
        title = "{Shock breakouts from red supergiants: analytical and numerical predictions}",
      journal = {\mnras},
         year = 2020,
        month = may,
       volume = {494},
       number = {3},
        pages = {3927-3936},
          doi = {10.1093/mnras/staa924},
archivePrefix = {arXiv},
       eprint = {2003.14097},
 primaryClass = {astro-ph.HE},
       adsurl = {https://ui.adsabs.harvard.edu/abs/2020MNRAS.494.3927K}
}

@ARTICLE{2008ApJ...683L.135C,
       author = {{Chevalier}, Roger A. and {Fransson}, Claes},
        title = "{Shock Breakout Emission from a Type Ib/c Supernova: XRT 080109/SN 2008D}",
      journal = {\apjl},
         year = 2008,
        month = aug,
       volume = {683},
       number = {2},
        pages = {L135},
          doi = {10.1086/591522},
archivePrefix = {arXiv},
       eprint = {0806.0371},
 primaryClass = {astro-ph},
       adsurl = {https://ui.adsabs.harvard.edu/abs/2008ApJ...683L.135C}
}

@ARTICLE{2008Sci...321.1185M,
       author = {{Mazzali}, Paolo A. and {Valenti}, Stefano and {Della Valle}, Massimo and {Chincarini}, Guido and {Sauer}, Daniel N. and {Benetti}, Stefano and {Pian}, Elena and {Piran}, Tsvi and {D'Elia}, Valerio and {Elias-Rosa}, Nancy and {Margutti}, Raffaella and {Pasotti}, Francesco and {Antonelli}, L. Angelo and {Bufano}, Filomena and {Campana}, Sergio and {Cappellaro}, Enrico and {Covino}, Stefano and {D'Avanzo}, Paolo and {Fiore}, Fabrizio and {Fugazza}, Dino and {Gilmozzi}, Roberto and {Hunter}, Deborah and {Maguire}, Kate and {Maiorano}, Elisabetta and {Marziani}, Paola and {Masetti}, Nicola and {Mirabel}, Felix and {Navasardyan}, Hripsime and {Nomoto}, Ken'ichi and {Palazzi}, Eliana and {Pastorello}, Andrea and {Panagia}, Nino and {Pellizza}, L.~J. and {Sari}, Re'em and {Smartt}, Stephen and {Tagliaferri}, Gianpiero and {Tanaka}, Masaomi and {Taubenberger}, Stefan and {Tominaga}, Nozomu and {Trundle}, Carrie and {Turatto}, Massimo},
        title = "{The Metamorphosis of Supernova SN 2008D/XRF 080109: A Link Between Supernovae and GRBs/Hypernovae}",
      journal = {Science},
         year = 2008,
        month = aug,
       volume = {321},
       number = {5893},
        pages = {1185},
          doi = {10.1126/science.1158088},
archivePrefix = {arXiv},
       eprint = {0807.1695},
 primaryClass = {astro-ph},
       adsurl = {https://ui.adsabs.harvard.edu/abs/2008Sci...321.1185M}
}

@ARTICLE{2009ApJ...692.1131T,
       author = {{Tanaka}, Masaomi and {Tominaga}, Nozomu and {Nomoto}, Ken'ichi and {Valenti}, S. and {Sahu}, D.~K. and {Minezaki}, T. and {Yoshii}, Y. and {Yoshida}, M. and {Anupama}, G.~C. and {Benetti}, S. and {Chincarini}, G. and {Della Valle}, M. and {Mazzali}, P.~A. and {Pian}, E.},
        title = "{Type Ib Supernova 2008D Associated With the Luminous X-Ray Transient 080109: An Energetic Explosion of a Massive Helium Star}",
      journal = {\apj},
         year = 2009,
        month = feb,
       volume = {692},
       number = {2},
        pages = {1131-1142},
          doi = {10.1088/0004-637X/692/2/1131},
archivePrefix = {arXiv},
       eprint = {0807.1674},
 primaryClass = {astro-ph},
       adsurl = {https://ui.adsabs.harvard.edu/abs/2009ApJ...692.1131T}
}

@ARTICLE{2026GCN.44075....1H,
       author = {{Huang}, Q.~J. and {Zou}, Z.-C. and {Li}, D.~Y. and {Mao}, X. and {Pan}, H.~W. and {Einstein Probe Team}},
        title = "{EP260321a: refined analysis of the EP-WXT and EP-FXT observations, implying a possible supernova shock breakout candidate}",
      journal = {GRB Coordinates Network},
         year = 2026,
        month = mar,
       volume = {44075},
        pages = {1},
       adsurl = {https://ui.adsabs.harvard.edu/abs/2026GCN.44075....1H}
}

@ARTICLE{2026GCN.44068....1H,
       author = {{Huang}, Q.~J. and {Zou}, Z.~C. and {Mao}, X. and {Li}, D.~Y. and {Pan}, H.~W. and {Einstein Probe Team}},
        title = "{EP260321a: Einstein Probe detection of an X-ray transient}",
      journal = {GRB Coordinates Network},
         year = 2026,
        month = mar,
       volume = {44068},
        pages = {1},
       adsurl = {https://ui.adsabs.harvard.edu/abs/2026GCN.44068....1H}
}
\bibliographystyle{aasjournalv7}



\end{document}